\documentclass[12pt]{article}
\usepackage{latexsym,graphicx}
\usepackage{subfigure}
\usepackage{amssymb}
\usepackage{amsmath}
\usepackage{amscd}
\usepackage{amsthm}
\usepackage{float}
\usepackage[left=2cm,top=2.5cm,right=2.5cm,bottom=1.5cm]{geometry}
\theoremstyle{definition}
\newtheorem{definition}{Definition}    
\newtheorem{theorem}{Theorem}[section]       
\begin{document}
\begin{center}
\large{\bf{Space-time singularities and the theory of retracts}} \\
\vspace{5mm}
\normalsize{Nasr Ahmed$^a$ and H. Rafat$^b$}\\
\vspace{5mm}
\small{\footnotesize$^{a}$ \textit{Astronomy Department, National Research Institute of Astronomy and Geophysics, Helwan, Cairo, Egypt}\footnote{nasr.ahmed@nriag.sci.eg}} \\
\small{\footnotesize$^{b}$\textit{Department of Mathematics, Faculty of Science, Tanta University, Tanta, Egypt.}}\\
\end{center}  
\begin{abstract}

In previous publications, we have started investigating some possible applications to the retraction theory in gravitational physics and showed that it can be very useful in providing proofs and explaining the topological bases. In the current work, we investigate the role of the retraction theory in the mathematical study of space-time singularities and suggest a topological restriction on the formation of singularities. Although we present a toy model using a 5D cosmological metric as an example, the current study applies for any collapsing system under the influence of gravity such as a massive star or the whole universe. After showing that the cosmological space-time $M$ can be retracted into lower dimensional circles $S_i \subset M $, we prove the existence of a homotopy between this retraction and the identity map which defines a deformation retract on $M$. Since a circle doesn't deformation retract onto a point but it does retract to a point, the defined deformation retract stops any circle $S_i$ from retracting into a point which means no singularity is formed. The paper underlines the role of algebraic topology in the mathematical study of space-time singularities and represents a new application of the retraction theory in mathematical physics.

\end{abstract}

\section{Introduction and motivation} \label{secintro}

The formation of space-time singularities due to gravitational collapse of massive objects is one of the most challenging aspects of General Relativity (GR). Gravitational collapse is also the main process responsible for cosmic structure formation where stars and galaxies form from interstellar gas under the effect of gravity. Depending on its mass, the final outcome of the gravitational collapse of a massive star which has
exhausted its nuclear fuel can be a white dwarf, a neutron star, or an Unhindered collapse to zero volume \cite{open}. To have a complete description of what happens after the horizon forms, we need a complete consistent theory of quantum gravity together with the appropriate calculational tools. Since such apparatus is not currently available, we can only study toy models where quantum contributions are considered \cite{open2}.\par

The mathematical definition of singularities in GR is given in terms of the geodesic incompletness \cite{g1,g2} in which a space-time is singular if there exists at least one incomplete geodesic, i.e. a geodesic that cannot be extended in at least one direction. According to singularity theorems, space-time singularities are inevitable and they represent a limit on the validity of the classical gravity theory. Consequently, it is widely accepted that a generalized quantum theory of gravity might be able to remove such singularities. Many candidates for this generalized quantum gravity theory have been suggested, with different proposals, such as Wheeler-DeWitt theory \cite{ww}, superstring theory \cite{ss}, brane theories \cite{bb}, loop quantum gravity \cite{loop2} and higher-order gravity \cite{hh}. Based on results of loop quantum cosmology, it is generally believed that loop quantum gravity can eliminate cosmological singularities such as the big bang and big crunch \cite{loop1, loop3}. Oscillatory cosmological models represent another way to overcome the initial singularity where the current Universe is a result of the collapse of a previous universe \cite{cy1,cy3,nascyc}. It has been also shown that a nonsingular Big Bounce appears naturally in the Einstein-Cartan-Sciama-Kibble theory of gravity \cite{pap}. Singularity-free cosmological solutions have also been obtained in string theories \cite{st}, quadratic gravity \cite{qa} and Einstein-Maxwell equations \cite{es}. A singularity-Free Cylindrical Cosmological Model has been given in \cite{ci}. So, it's highly important to get a deeper understanding to the formation of space-time singularities due to gravitational collapse. \par

The current paper discusses the formation of space-time singularities at the end of the gravitational collapse process from a global topological view point. We introduce a toy model with a cosmological metric as an example, and we are going to discuss the possible topological restrictions on the formation of a space-time singularity as a result of the gravitational collapse. A 'singularity' here means the collapsing into a single 'point' of zero dimension. To achieve this task, we make use of the topological theory of retracts and the homotopy theory. Topology is the ideal tool for the study of shapes and spaces which can be continuously deformed into each other. Space-time in GR exists as a manifold, and that makes the use of topological notions and methods in GR possible. Applications of differential topology in GR has been discussed in \cite{diffg}, and the topology change in GR has been investigated in \cite{topc}. The topological quantum field theory (TQFT) was introduced in \cite{tqft}. \par

Homotopy can be defined as a continuous deformation of one continuous function to another \cite{kans}, and it is used to describe deformations in topology. A nice example is human aging which is a continuous process where the topological shape of a child is related to the shape of an old person by a homotopy describing the shape at every age between \cite{kans}. Thus, homotopy is an equivalence relation between maps and and it has been applied to many areas in mathematical physics \cite{tqft,cond, cond2, ho, hoo, hooo}. In condensed matter physics, it has been shown that the use of the homotopy theory can resolve all the paradoxes appear from the old theory of defects in ordered media \cite{cond, cond2}. A particular class of defects that can not be removed by continuous deformation of fields is the topological defects which appear in both condensed matter physics and cosmology. This particular class of defects is topologically distinct solutions to nonlinear partial differential equations where the homotopy theory can be used in determining when the solutions are truly distinct. In other words, we have a homotopical distinction between the vacuum solution and the solution in which these objects appear. the homotopy quantum field theory is a branch of the topological quantum field theory \cite{tqft} in which theoretical physics' ideas are applied to study homotopy classes of maps from manifolds to a fixed target space. The important role of the homotopy theory in constructing a theory of quantum gravity has been presented in \cite{hoo} where a finite-dimensional quantum theory has been built from GR by a homotopy method. The role of homotopy in algebraic quantum field theories has been illustrated in \cite{hooo}. \par

While the homotopy theory has been applied to several areas in theoretical physics, most of the studies on the topological retraction theory -if not all- are pure mathematical studies. In previous publications \cite{nas1, nas2, nas3}, some useful applications of the retraction theory in quantum gravity have been introduced. The retraction, deformation retract and folding of the 5D Schwarzchid field have been discussed in \cite{nas1}. In \cite{nas2}, the connection between the retraction theory and physical holography has been investigated and a mathematical restatement to the holographic principle has been introduced. In particular, we have shown that the theory of retracts provides a solid topological base to the holographic principle and can be used to explore the geometry of the hologram boundary. In \cite{nas3}, we found that applying the retraction theory to wormholes and black holes implies that these objects can get continuously deformed and reduced to lower dimensions. We have also indicated that while homotopy provides a topological base to wormholes/compact objects' deformations, the retraction provides a topological base to the dimensional reduction which is expected to take place near the Planck length where the effective space-time dimension reduces to lower dimensions \cite{viii}. Thus, the retraction theory has provided a rigorous proof to the existence of deformations in black holes/wormholes and explained the topological origin of such deformations. \par

The retraction method we are going to use in the current work has been developed by the authors and used in \cite{nas1, nas2, nas3}. Although in this paper we study a toy model with a Ricci-flat cosmological metric, the retraction method can be extended and applied to non-Ricci-flat geometries. The main physical motivations behind discussing a collapsing Ricci-flat universe in our toy model are: (1) The strong analogy between a gravitationally collapsing system and the topological retraction. On the mathematics side we have a continuously shrinking space $X$ into a subspace $A\subset X$, while on the physics side we have a continuously collapsing (contracting) space-time volume (including matter and fields). (2) Studying Ricci-flat manifolds have been an important topic both in physics and geometry \cite{zu}. In GR, they represent vacuum solutions to Einstein equations for Riemannian manifolds with zero cosmological constant (Ricci-flat manifolds are Riemannian manifolds with vanishing Ricci curvature). A particular case of Ricci-flat manifolds are the well-known Calabi-Yau manifolds which have many applications in theoretical physics, specially in superstring theory where the 6 extra dimensions can be compactified on a Calabi-Yau manifold. The retraction technique we use here is based on comparing the space-time metric with the general form of the flat Minkowiskian metric $ds^{2}= -dx_{o}^{2}+ \sum_{i=1}^4 dx_{i}^{2}$ and obtaining a set of coordinate transformations with a set of integration constants. Then, the role of Euler-Lagrange equations comes to study the geodesics as we will in section (\ref{sec3}). \par

Let’s see how the current study is related to the previous studies on space-time singularities. First, the current study discusses singularities from a global topological side while previous studies concentrated on the local geometrical side. Strictly speaking, the role of algebraic topology in the mathematical study of space-time singularities has not been investigated before. Secondly, since a complete consistent theory of quantum gravity which provides a complete understanding to singularities still never been reached yet, using solid topological notions for the description of singularities might give a deeper understanding. The current work suggests a mathematical restriction on the existence of space-time singularities using topological notions. Thirdly, the paper introduces a new application of the retraction theory in mathematical physics. We believe that this topological theory is so powerful in explaining some modern concepts of theoretical physics. 

The paper is organized as follows: The introduction and motivation behind the current work is included in section 1. The mathematical definitions of the topological concepts used in the current work are included in section 2. The retraction of the cosmological space-time $M$ into circles $S_i \subset M$ is performed in section 3. In section 4, we prove the existence of a deformation retract on $M$ and discuss the restriction on the formation of the singularity. The conclusion is included in section 5.

\section{Basic definitions} \label{sec1}
In this section we introduce the basic mathematical definitions with a summary to the work plan in the next sections.
\theoremstyle{definition}
\begin{definition}
A subspace $A$ of a topological space $X$ is called a retract of $X$, if there exists a continuous map $r : X \rightarrow A $ such that $X$ is open and $r(a) = a$ (identity map), $\forall a \in A$. Because the continuous map $r$ is an identity map from $X$ into $A \subset X$, it preserves the position of all points in $A$. A good example is the retraction $r:S^2 \rightarrow \left\{(1,0,0)\right\} \in S^2$ that takes the sphere to a point defined by $r(x,y,z)=(1,0,0)$ \cite{kans}.
\end{definition}
\theoremstyle{definition} \label{def2}
\begin{definition}{Homotopy:} \cite{geo} Let $X$, $Y$ be smooth manifolds and $f: X \rightarrow Y$ a smooth map between them. A homotopy or deformation of the map $f$ is a smooth map $F: X \times [0,1] \rightarrow Y$ with the property $F(x,0)=f(x)$. Each of the maps $f_t(x)=F(x,t),~t\in [0,1]$ is said to be homotopic to the initial map $f_0=f$ and the map of the whole cylinder $X \times [0,1]$ is called a homotopy.
Two continuous maps $f: X \rightarrow Y$ and $g: X \rightarrow Y$ are called homotopic if there exists a continuous map $F: X \times [0,1] \rightarrow Y$ with $F(x,0)=f(x)$ and $F(x,1)=g(x)$ $\forall x \in X$.
\end{definition}
\theoremstyle{definition}
\begin{definition}{Deformation retract:}\label{def1}
A subset $A$ of a topological space $X$ is said to be a deformation retract if there exists a
retraction $r: X \rightarrow A$, and a homotopy $f : X \times [0,1] \rightarrow X $ such that \cite{11}: $f (x, 0) = x$ $\forall x\in X$, $f(x, 1) = r (x)$ $\forall r\in X$, $f(a, t) = a$ $\forall a\in A, t \in [0,1]$.
\end{definition}
\begin{definition}{contractible space:} \cite{course,hatcher}\label{def4}
A space $X$ is said to be contractible if there is some point $p \in X $ so that $p$ is a deformation retract of $X$ ( i.e., if it is homotopy equivalent to a point). Intuitively, it is a space that can be contracted to a point.
\end{definition}
Let's take the circle as an example on definition \ref{def4}. A circle $S$ cannot be deformed within itself to a point \cite{course}, which means that the circle is not contractible (a circle can not deformation retract into a point). This also comes from the fact that the fundamental group of a circle $\pi_1(S^1)=\mathbb{Z}$ while all the homotopy groups of a contractible space are trivial. The fundamental group of a point $p$ is zero, $\pi_1(p)=0$. In other words, a point cannot be the deformation retract of a circle or any other space with a nontrivial homotopy group. However, A circle $S^1$ can retract onto a point $p$ where this retraction is just a constant map $r:S^1 \rightarrow p$.\par 
The fact that:  \textit{A circle does not deformation retract onto a point but it does retract to a point} represents the central idea of the current work which can be summarized as follows: After we perform a retraction of the $5D$ cosmological space-time $M$ onto $4D$ circles $S_i \subset M$, these circles $S_i$ can now be retracted to a point. The physical analogy of this space-time shrinking into a point is the formation of a space-time singularity where the whole universe collapses into a point. We then proceed to prove the existence of a deformation retract on $M$ by defining the appropriate homotopy. But as we stated above, a circle does not deformation retract onto a point which means that the circles $S_i \subset M$ can not continue shrinking into a point. The physical meaning of this situation is that the space-time singularity ( defined as a dimensionless point) can not be reached, i.e. there will be no singularity due to the existence of homotopy which defines a deformation retract on $M$.

\section{Cosmological solution and its retraction.} \label{sec3}

In this section we perform a retraction to the cosmological space-time $M$ into subspaces. A new class of solutions of the 5D Einstein equations which represents an extension to the FRW solutions is given by \cite{basic1}:
\begin{eqnarray} \label{mett}
ds^{2}&=& B^2 dt^2-A^2 \left(\frac{dr^2}{1-\kappa r^2}+r^2 d \Omega^2 \right)-dy^2\\  \nonumber
A^2&=& (\mu^2-\kappa)y^2+2\nu y+\frac{\nu^2+K}{\mu^2+\kappa}\\   \nonumber
B&=&\frac{1}{\mu}\frac{\partial A}{\partial t} \equiv \frac{\dot{A}}{\mu}
\end{eqnarray}
Where $\mu=\mu(t)$ and $\nu=\nu(t)$ are arbitrary functions, $\kappa$ is the 3D curvature index ($k = \pm 1$, $0$)
and $K$ is a constant. It has been concluded that the 5D models (\ref{mett}) may have two kinds of singularities, big bang at $A=0$ and big bounce at $B=0$. Since recent observations suggest that the present universe is flat \cite{ob1, ob2}, we set $\kappa =0$. As we have indicated in the introduction, the reasons why we have chosen the Ricci-flat cosmological metric (\ref{mett}) for our toy model are: 1- it represents an extension to the FRW solutions. 2- the special importance of Ricci-flat manifolds in both gravitational physics and geometry \cite{zu}.  Without loss of generality, the solution (\ref{mett}) can be written as \cite{basic1}:
\begin{eqnarray} \label{mett2}
ds^{2}&=& B^2 dt^2-A^2 \left(dr^2+r^2 d \Omega^2 \right)-dy^2\\  \nonumber
A^2&=& t^{2n}(y-Ct)^2+Kt^{-2n}\\   \nonumber
B&=& \frac{\dot{A}}{t^n}
\end{eqnarray}
Comparing the general 5D flat metric 
\begin{equation}
ds^{2}= -dx_{o}^{2}+ \sum_{i=1}^4 dx_{i}^{2}   \label{flatt}
\end{equation}
with the metric (\ref{mett2}) for $K=0$ (for simplicity) and making use of the basic metric definition $ds^2 = g_{ij} dx^i dx^j$, we obtain the following class of coordinate transformations
\begin{eqnarray} \label{xi}
x_o&=&\pm \left(\frac{2n^3(C-y)^2+n^2(5C^2-2Cy-3y^2)+4n(C+\frac{1}{2}y)^2+C(C+2y)}{2t^{2n}(4n^3-4n^2-n+1)}+C_o\right)^{\frac{1}{2}},\\  \nonumber
x_1&=&\pm \sqrt{Ar^2+C_1}, \  \  \  x_2=\pm \sqrt{A^2r^2\theta^2+C_2}, \  \  \ x_3=\pm \sqrt{A^2r^2 \sin^2\theta \phi^2+C_3},\\  \nonumber
x_4&=&\pm \sqrt{\frac{1}{2}y^2+C_4}.
\end{eqnarray}
Where $C_{o}$, $C_{1}$, $C_{2}$, $C_{3}$ and $C_{4}$ are constants of integration. We are going to make use of the coordinate relations (\ref{xi}) in performing the geodesic retractions of the $5$D cosmological space-time $M$ given by (\ref{mett2}). To find a geodesic which is a subset of $M$, we can use the geodesic equation or the Euler-Lagrange equations where they both equivalent in GR. The Euler-Lagrange equations associated to the Lagrangian $L(x^{\mu},\dot{x}^{\mu})=\frac{1}{2}g_{\mu \nu}\dot{x}^{\mu}\dot{x}^{\nu}$ are
\begin{equation}
\frac{d}{d\lambda}\left(\frac{\partial L}{\partial \dot{x}^{\alpha}}\right)-\frac{\partial L}{\partial x^{\alpha}} =0, ~~~~i=1,2,3,4
\end{equation}
We start from the Lagrangian
\begin{equation}
L=\frac{\dot{A}^2}{2t^{2n}}\dot{t}^2-\frac{1}{2}A^2(\dot{r}^2+r^2\dot{\theta}^2+r^2 \dot{\phi}^2\sin^2\theta )-\frac{1}{2}\dot{y}^2. \label{Lagr}
\end{equation}
No explicit dependence on either $\phi$, so $\frac{\partial L}{\partial \dot{\phi}}$ is a constant of motion, i.e.
\begin{equation}\label{2}
-t^{2n}(y-Ct)^2 r^2\sin^2\theta \, \dot{\phi}=h.
\end{equation}
The $r$, $\theta$, $\phi$, $t$ and $y$-components respectively give
\begin{equation}\label{3}
\frac{d}{d\lambda}(A^2\dot{r})-A^2r(\dot{\theta}^2+\sin^2\theta \, \dot{\phi}^2)=0
\end{equation}
\begin{equation}
\frac{d}{d\lambda}(A^2r^2\dot{\theta})-A^2r^2 \sin \theta \cos \theta \, \dot{\phi}^2=0
\end{equation}
\begin{equation}
\frac{d}{d\lambda}\left[t^{2n}(y-Ct)^2 r^2\sin^2\theta \, \dot{\phi}\right]=0.
\end{equation}
\begin{equation}
\frac{d}{d\lambda}\left[\left(Ct(n+1)-ny\right)^2\dot{t}t^{-2n-2}\right]-\frac{1}{2}\left[\dot{t}^2f(t)-g(t)(\dot{r}^2+r^2\dot{\theta}^2+r^2\sin^2 \theta \dot{\phi}^2)\right]=0
\end{equation}
where $g(t)$ and $f(t)$ are given as
\begin{equation}
g(t)=\frac{\partial A^2}{\partial t}=2t^{2n}(Ct-y)(C+nt^{-1}(Ct-y))
\end{equation}
\begin{eqnarray}
f(t)&=&\frac{\left(2nt^{2n-1}(y-Ct)^2-2Ct^{2n}(y-Ct)\right)^2}{2t^{6n}(y-Ct)^2}\left(\frac{C}{y-Ct}-\frac{3n}{t}\right)\\ \nonumber
&+&\frac{2nt(y-Ct)-2C}{2t^{2n+2}(y-Ct)}\left((y-Ct)^2(4n^2-2n)-8ntC(y-Ct)+2C^2t^2\right).
\end{eqnarray}
\begin{equation}
\frac{d}{d\lambda}\dot{y}+\frac{1}{2}\left[\dot{t}^2f_1(t)+(\dot{r}^2+r^2\dot{\theta}^2+r^2\sin^2 \theta \dot{\phi}^2)g_1(t)\right]=0
\end{equation}
where $g_1(t)$ and $f_1(t)$ are given as
\begin{equation}
g_1(t)=\frac{\partial A^2}{\partial y}=2t^{2n}(Ct-y)
\end{equation}
\begin{equation}
f_1(t)=-\frac{\left(2nt^{2n-1}(y-Ct)^2-2Ct^{2n}(y-Ct)\right)^2}{2t^{6n}(y-Ct)^3}+\frac{nt^{-1}(y-Ct)^2-C(y-Ct)}{t^{2n}(y-Ct)^2}.
\end{equation}
Although we can not study the retractions for $t=0$, we are still able to study some retractions of $M$ and see what types of geodesic retractions we get. From (\ref{2}), if $h=0$ then $\phi=H$ where $H$ is a constant, or $\theta=0$. For $H=0$, the coordinates (\ref{xi}) become
\begin{eqnarray} \label{coos1}
x_o^{\phi=0}&=&\pm \left(\frac{2n^3(C-y)^2+n^2(5C^2-2Cy-3y^2)+4n(C+\frac{1}{2}y)^2+C(C+2y)}{2t^{2n}(4n^3-4n^2-n+1)}+C_o\right)^{\frac{1}{2}},\\  \nonumber
x_1^{\phi=0}&=&\pm \sqrt{Ar^2+C_1}, \  \  \  x_2^{\phi=0}=\pm \sqrt{A^2r^2\theta^2+C_2}, \  \  \ x_3^{\phi=0}=\pm \sqrt{C_3}, \  \  \ x_4^{\phi=0}=\pm \sqrt{\frac{1}{2}y^2+C_4}.
\end{eqnarray}
Since $ds^2=x_{1}^{2}+x_{2}^{2}+x_{3}^{2}-x_{o}^{2}>0$ which is a circle $S_1 \subset M$. This geodesic is a retraction in the 5D space-time $M$ represented by the metric (\ref{mett2}). For $\theta=0$, we get
\begin{eqnarray} \label{coos2}
x_o^{\theta=0}&=&\pm \left(\frac{2n^3(C-y)^2+n^2(5C^2-2Cy-3y^2)+4n(C+\frac{1}{2}y)^2+C(C+2y)}{2t^{2n}(4n^3-4n^2-n+1)}+C_o\right)^{\frac{1}{2}},\\  \nonumber
x_1^{\theta=0}&=&\pm \sqrt{Ar^2+C_1}, \  \  \  x_2^{\theta=0}=\pm \sqrt{C_2}, \  \  \ x_3^{\theta=0}=\pm \sqrt{C_3}, \  \  \ x_4^{\theta=0}=\pm \sqrt{\frac{1}{2}y^2+C_4}.
\end{eqnarray}
Since $ds^2=x_{1}^{2}+x_{2}^{2}+x_{3}^{2}-x_{o}^{2}>0$ which is a circle $S_2 \subset M$. This geodesic is a retraction in the 5D space-time $M$ represented by the metric (\ref{mett2}). So, we have a retraction of $M$ defined as $R : M \Rightarrow S_i ,\,i=1,2$, and the following theorem has been proved
\begin{theorem}
Some types of the geodesic retractions of the 5D cosmological space-time $M$ are circles $S_i \subset M$. 
\end{theorem}
\section{Homotopy and deformation retract}
In the previous section, we have shown that the space-time $M$ can get retracted into lower dimensional circles $S_i$. Recalling section 2, a circle $S$ can get retracted onto a point (singularity) unless there is a homotopy defines a deformation retract on $M$. In this case, with a deformation retract defined on $M$, a circle $S$ can not get deformation retracted onto a point which means no formation of a space-time singularity. \par
Recalling definition (\ref{def1}), deformation retract is a homotopy between the retraction and the identity map. So, in addition to the retraction $R: M \rightarrow S_i$, a homotopy must be defined between the retraction and the identity map on $M$ ( $f : M \times [0,1] \rightarrow M $ ) such that $f (s, 0) = s$ $\forall s\in M$, $f(s, 1) = R (s)$ $\forall R\in M$, $f(a, t) = a$ $\forall a\in S_i, t \in [0,1]$ to have a deformation retract on $M$. For the retraction of $M$ into a geodesic $S_1 \subset M$, a homotopy can be defined as $\xi^{S_1} : M \times [0,1] \rightarrow M$ where
\begin{equation} \label{hom2}
\xi^{S_1}(m,s)=\cos \frac{\pi s}{2}~\xi(m,0)+\sin \frac{\pi s}{2}~\xi^{S_1}(m,1),~~~~~~~~~~~~\forall m \in M~ \&~ \forall s \in [0,1].
\end{equation}
With
\begin{equation}
\xi(m,0)=\left\{x_o, x_1, x_2, x_3, x_4\right\}~~\& ~~\xi^{S_1}(m,1)=\left\{x_o^{\phi=0}, x_1^{\phi=0}, x_2^{\phi=0}, x_3^{\phi=0}, x_4^{\phi=0}\right\}.
\end{equation}
And the coordinates $x_i^{\phi=0},~i=0, 1, 2, 3, 4$ are given by (\ref{coos1}) . The homotopy of the retraction of $M$ into a geodesic $S_2 \subset M$, is defined as
\begin{equation} \label{hom1}
\xi^{S_1}(m,s)=(1-s)~\xi(m,0)+s~\xi^{S_1}(m,1),~~~~~~~~~~~~\forall m \in M~ \&~ \forall s \in [0,1].
\end{equation}
With
\begin{equation}
\xi(m,0)=\left\{x_o, x_1, x_2, x_3, x_4\right\}~~\& ~~\xi^{S_1}(m,1)=\left\{x_o^{\theta=0}, x_1^{\theta=0}, x_2^{\theta=0}, x_3^{\theta=0}, x_4^{\theta=0}\right\}.
\end{equation}
And the coordinates $x_i^{\theta=0},~i=0, 1, 2, 3, 4$ are given by (\ref{coos2}). With the defined homotopies, we now have a deformation retract defined on $M$. \par
For the sake of completeness, we carefully discuss the basic mathematical aspects of the defined homotopies such as existence, uniqness/classes of transformations and well-definiteness. The geodesic retractions we have found are all circles $S_i \in W$. For a homotopy $\xi^{S_i} : M \times [0,1] \rightarrow M$ to be unique, it must have the same form $\forall i$. However, this is not the case as the two homotopies (\ref{hom2}) for $i=1$ and (\ref{hom1}) for $i=2$ are different. We have replaced $\cos \frac{\pi s}{2}$ and $\sin \frac{\pi s}{2}$ in $(\ref{hom2})$ with $(1-s)$ and $s$ in (\ref{hom1}) so they all satisfy definition (\ref{def1}). Then, the defined homotopy between the retraction and the identity map on $M$ is not unique which can also be directly deduced from the fact that a space can deformation retract into different subspaces. The homotopy $\xi^{S_i} : M \times [0,1] \rightarrow M$ is well defined and does exist under the three conditions in definition (\ref{def1}). Any other possible homotopies are expected to lead to similar results as they must satisfy definition (\ref{def1}).\par
\section{Conclusion}
The possible role of algebraic topology in the mathematical study of space-time singularities has been investigated through the theory of retracts. Since a complete consistent theory of quantum gravity has not been discovered yet, our understanding to space-time singularities is not complete. The incomplete understanding of space-time singularities is one of the main motivations behind the current study, using rigorous topological notions can lead to a deeper understanding to this challenging subject. We have presented a toy model with a Ricci-flat cosmological metric as an example. However, the study applies for any collapsing (contracting) system under the influence of gravity such as a massive star or the whole universe. We have shown that the space-time $M$ can get retracted into circles $S_i$, then we proved the existence of a deformation retract defined on $M$. The defined deformation retract stops any circle $S_i$ from retracting into a point which means no formation of space-time singularity. Some basic mathematical aspects of the defined homotopies such as existence, unicity/classes of transformations and well definiteness have been discussed. The paper represents a new application of the retraction theory in mathematical physics.
\section*{Acknowledgment}
We are so grateful to the reviewer for his many valuable suggestions and comments that significantly improved the paper.

\end{document}